\documentclass[aps,prd,preprint,superscriptaddress,tightenlines,nofootinbib]{revtex4}

\usepackage{graphicx}% Include figure files
\usepackage{dcolumn}% Align table columns on decimal point
\usepackage{bm}% bold math
\usepackage{epsfig}
\usepackage[dvips]{color}
\usepackage{graphics,color,epsfig}
\newcommand{\bodyspacing}{1}
\newcommand{\bbsp}{\renewcommand{\baselinestretch}{\bodyspacing}\small\normalsize}
\newcommand{\ebsp}{\par\renewcommand{\baselinestretch}{1}\small\normalsize}

\newcommand{\bfigenvh}{\ebsp\begin{figure}[h]}
\newcommand{\bfigenvhh}{\ebsp\begin{figure}[h!]}
\newcommand{\bfigenv}{\ebsp\begin{figure}}
\newcommand{\efigenv}{\end{figure}\bbsp}
\newcommand{\btabenvh}{\ebsp\begin{table}[h]}
\newcommand{\btabenv}{\ebsp\begin{table}}
\newcommand{\etabenv}{\end{table}\bbsp}
\newcommand{\bi}{\ebsp\begin{itemize}}
\newcommand{\ei}{\end{itemize}\bbsp}
\newcommand{\dppp}{$D^0 \rightarrow \pi^- \pi^+ \pi^0$}
\newcommand{\dbppp}{$\overline{D^0} \rightarrow \pi^+ \pi^- \pi^0$}
\newcommand{\dkpp} {$D^0 \rightarrow K^- \pi^+ \pi^0$}

\newcommand{\ks}{$K^0_{S}$ }

\newcommand{\acp}{${\mathcal A}_{CP}$ }
\newcommand{\red}[1]{\textcolor[rgb]{0.5,0,0}{#1}}

\newcommand{\rightLarrow} 
{\mbox{
	\begin{picture}(0,10)(0,0)
	\put(0,10){\line(0,-1){5}}
	\put(1,5){\oval(2,2)[bl]}
	\put(1.5,-0.1){$\rightarrow$}
	\put(5,3){~}
	\end{picture}
\hspace*{-0.115in}
~~
}}

\newcommand{\bc}{\begin{center}}
\newcommand{\ec}{\end{center}}
\newcommand{\be}{\begin{equation}}
\newcommand{\ee}{\end{equation}}

\def\mtiny{\vrule width 0pt}
\def\mrm#1{\mathrm{#1}}
\def\DZ{\relax\ifmmode{D^0}\else{$\mrm{D}^{\mrm{0}}$}\fi}
\def\Dz{\relax\ifmmode{D^0}\else{$\mrm{D}^{\mrm{0}}$}\fi}
\def\DZB{\relax\ifmmode{\overline{D}\mtiny^0}
        \else{$\overline{\mrm{D}}\mtiny^{\mrm{0}}$}\fi}
\def\Dzb{\relax\ifmmode{\overline{D}\mtiny^0}
        \else{$\overline{\mrm{D}}\mtiny^{\mrm{0}}$}\fi}
\def\KZ{\relax\ifmmode{K^0}\else{$\mrm{K}^{\mrm{0}}$}\fi}
\def\KZB{\relax\ifmmode{\overline{K}\mtiny^0}
        \else{$\overline{\mrm{K}}\mtiny^{\mrm{0}}$}\fi}
\def\BZ{\relax\ifmmode{B^0}\else{$\mrm{B}^{\mrm{0}}$}\fi}
\def\BZB{\relax\ifmmode{\overline{B}\mtiny^0}
        \else{$\overline{\mrm{B}}\mtiny^{\mrm{0}}$}\fi}
\def\DZS{\relax\ifmmode{D^{*+}}\else{$\mrm{D}^{\mrm{*+}}$}\fi}
\def\DZM{\relax\ifmmode{D^{*-}}\else{$\mrm{D}^{\mrm{*-}}$}\fi}
\def\DZC{\relax\ifmmode{\overline{D}\mtiny^0}
        \else{$\overline{\mrm{D}}\mtiny^{\mrm{0}}$}\fi}
\def\DZD{\relax\ifmmode{{\bar D}\mtiny^0}
        \else{${\bar{\mrm{D}}}\mtiny^{\mrm{0}}$}\fi}

\begin{document}

%\preprint line(s) will be ignored for PRL/PRD
%\preprint{CLEO Draft 03-01A} % For paper draft CBX YY-NN -> Draft YY-NNA
\preprint{CLEO CONF 03-03}   % For conference papers
\preprint{EPS-208}      % For conference papers
%\preprint{CLEO 03-02}         % for CLNS notes

\title{ A Dalitz Plot Analysis of $D^0
\rightarrow \pi^- \pi^+ \pi^0$ Decays in CLEO II.V}
% for conference papers (ask CLEOAC for appropriate text)
\thanks{Submitted to the 
International Europhysics Conference on High Energy Physics,
July 2003, Aachen}

%-------- INSERT HERE ------------
\author{V.~V.~Frolov}
\author{K.~Y.~Gao}
\author{D.~T.~Gong}
\author{Y.~Kubota}
\author{S.~Z.~Li}
\author{R.~Poling}
\author{A.~W.~Scott}
\author{A.~Smith}
\author{C.~J.~Stepaniak}
\author{J.~Urheim}
\affiliation{University of Minnesota, Minneapolis, Minnesota 55455}
% ...
%\author{(CLEO Collaboration)} %FOR PRD_SPECIAL_CHANGEME
\collaboration{CLEO Collaboration} %FOR PRL,CLNS
\noaffiliation
%-------- END INSERT ------------

%please hard code the date when you have a final draft and submit to CLEOAC
\date{\today}

\begin{abstract} 
Using the $9 fb^{-1}$ data sample collected with the CLEO
II.V detector at the Cornell Electron Storage Ring, we have studied the
resonant substructure of the Cabibbo suppressed decay $D^0 \rightarrow
\pi^-\pi^+\pi^0$.  We observe significant contributions from the
$\rho^-\pi^+$, $\rho^+\pi^-$, $\rho^0\pi^0$, and non-resonant
channels, and present preliminary results of the amplitudes, phases,
and fit fractions for these sub-modes. No evidence for the
$\sigma(500)$ or more massive $\rho$ resonances was found. We observe no
CP violation, finding \acp = $0.01^{+0.09}_{-0.07} \pm 0.09 $.  
\end{abstract}

\maketitle

% Insert body of the text here.
% Hard-wired - RSG - 3Jun03

\section{Introduction}
\subsection{Motivation}\label{subsec:prod}

Three-body decays provide a rich laboratory in which to study the interference 
between intermediate state resonances, and can provide a direct probe of final 
state interactions. When a particle decays into three or 
more daughters, intermediate resonances dominate the decay rate and
will cause a non-uniform distribution of events in phase space 
when analyzed using a ``Dalitz plot'' technique.\cite{dal} 
Since all events of a particular decay mode have the same final state, 
multiple resonances at the same location in phase space will interfere.  
This provides the opportunity to experimentally measure both the amplitudes 
and phases of the intermediate decay channels, which in turn allows us to calculate 
their fit fractions as well as set limits on CP violating asymmetries.
 
This paper describes a study of the underlying structure in the decay 
$D^0 \rightarrow \pi^-\pi^+\pi^0$, and is motivated in part by E791's 
recent work on $D^+ \rightarrow \pi^- \pi^+ \pi^+$ which found
significant evidence for a broad neutral scalar resonance, 
the $\sigma(500)$.\cite{e791}

Since \dppp~is a CP eigenstate, we also look for CP
violation.  With no CP violation, $D^0 \rightarrow \rho^+ \pi^-$
should have the same amplitudes and phases as $\overline{D^0}
\rightarrow \rho^-\pi^+$.  Recent theoretical works suggest that CP
violation in \dppp~ may be as large as $0.1\%$.
\cite{Buccella,Santorelli}.

\subsection{Three Body Decays}
In this analysis we are studying the decay of a spin zero $D$ into
three spin zero $\pi$'s, hence two degrees of freedom are needed to completely 
describe the system.
\footnote{Each daughter has 4 degrees of freedom, for a total of 12.
Energy and momentum conservation fixes 4 of these. Knowledge of the daughter masses 
fixes 3 more. Since overall spatial orientation is irrelevant,
3 more are fixed, leaving 2.}
 A convenient choice is to pick any two of the $\pi \pi$ invariant
mass squared terms, for example $m^2_{\pi^- \pi^+}$ and $m^2_{\pi^+ \pi^0}$ since, when
averaged over intermediate spin states, the phase-space for the decay is 
independent of position in ($m^2_{\pi^-\pi^+}, m^2_{\pi^+\pi^0}$) space.
A Dalitz plot is simply a scatter plot of all event candidates in the
($m^2_{\pi^- \pi^+}$, $m^2_{\pi^+ \pi^0}$) plane.  
Since phase-space is uniform in these variables, any structure that shows
up in the Dalitz plot is due entirely to the matrix element of the decay. 
Intermediate resonances will show up as bands on the Dalitz plot. 
Loosely stated, the structure within each band tells us about the spin of the resonance, 
the number of events in each tells us about relative amplitudes, and the regions 
of overlap between the bands contains information about relative phases.

\begin{figure}[htbp]
\centerline{\epsfxsize=4.5in\epsfbox{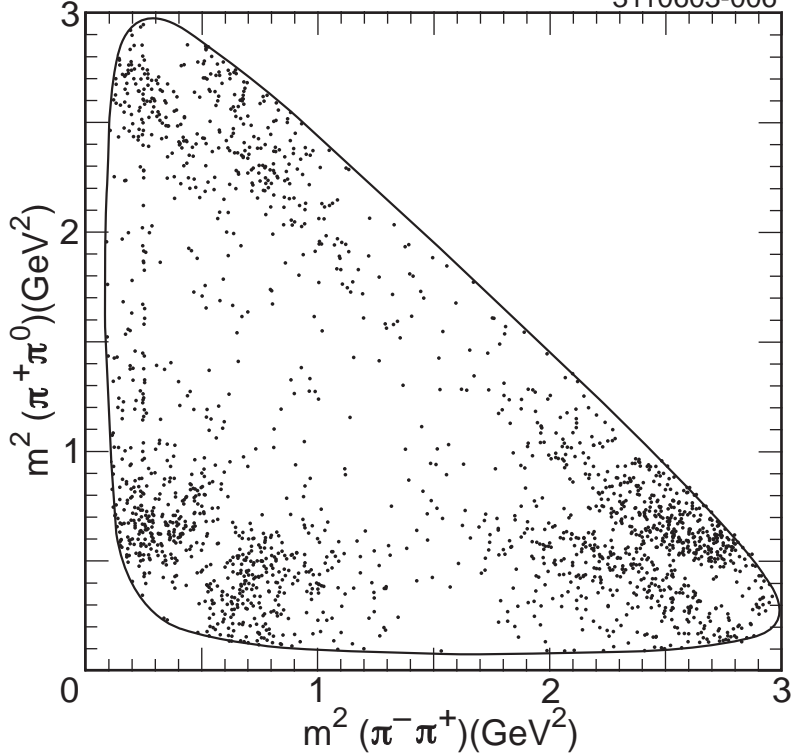}}
\caption{ \label{fig:dalitz}The $D^0 \rightarrow
\pi^- \pi^+ \pi^0$ Dalitz plot after final event selection.}
\end{figure}

Figure~\ref{fig:dalitz} shows a Dalitz plot
of all events passing the analysis selection requirements in this analysis. 
It is clear that the distribution of events is not uniform. Indeed, bands centered at 
mass-squared values appropriate for $\rho(770)$ mesons can clearly be seen along both
axes. The fact that these bands are not uniform, but rather are populated toward the 
edges of the plot, indicates that they are due to spin-1 resonances.  
The narrow uniform band at $m^2_{\pi^- \pi^+}\sim 0.25$ GeV$^2$ is due to 
$D^0 \rightarrow K^0_S \pi^0$ decays.

\subsection{Resonances}
The technique used in this analysis is to fit the ($m^2_{\pi^- \pi^+}$, $m^2_{\pi^+ \pi^0}$)
event distribution from data with various models containing collections of possible 
resonant as well as non-resonant contributions.  Any non-resonant contribution is 
assumed to be uniform across the Dalitz plot, and all resonances are represented 
by a Breit-Wigner lineshape times appropriate factors:

\be
\label{eqn:MDab2}
{\mathcal{M}}_{D^0 \rightarrow (AB)C}
= \\
F_{D^0}(q^2) F_{res}(q^2)
{ \begin{Huge}
 {A_S}
 \over
 {m^2_{res} - q^2 - im_{res} \Gamma(q)}
  \end{Huge}
},
\\
\ee

\noindent
where $q^2 = m^2_{AB}$ is the reconstructed invariant mass squared of the ($AB$) candidate, 
$m_{res}$ is the nominal mass of the resonance,
$F_{D^0}(q^2)$ and $F_{res}(q^2)$ are form-factors\cite{blatt}, 
and $\Gamma(q)$ is a mass-dependent width.\cite{pilkuhn} 
The term in the numerator, $A_S$, comes from
angular momentum considerations and depends on the spin $S$ of the resonance. 
For spin 0 resonances, $A_0$ = 1, and for spin 1, 
$A_1 = m^2_{AC} - m^2_{BC} + (m^2_{D^0}-m^2_C)(m^2_B-m^2_A)/m^2_{res}$.

\smallskip
It is unknown, a priori, how much of each possible resonance will
be required to fit the data, thus we
weigh each Breit-Wigner term in the matrix element with its own 
amplitude $\mathcal{A}$ and phase $\phi$. 
The total matrix element is simply a sum of the matrix elements for 
all resonances under consideration in a given model:

\be
\label{eqn:model}
\begin{array}{rl}
    \mathcal{M}_{D^0 \rightarrow \pi^- \pi^+ \pi^0} =
    & \mathcal{A}_{non~res} \times e^{i~\phi_{non~res}} ~~+ \\

    & \mathcal{A}_{\rho^+ \pi^-} \times e^{i~\phi_{\rho^+ \pi^-} } \times
      \mathcal{M}_{D^0 \rightarrow \rho^+ \pi^- } ~~+ \\

    & \mathcal{A}_{\rho^- \pi^+} \times e^{i~\phi_{\rho^- \pi^+} } \times
      \mathcal{M}_{D^0 \rightarrow \rho^- \pi^+ } ~~+ \\

    & \mathcal{A}_{\rho^0 \pi^0} \times e^{i~\phi_{\rho^0 \pi^0} } \times
      \mathcal{M}_{D^0 \rightarrow \rho^0 \pi^0 } ~~+ \ldots  \\
\end{array}
\ee

It is clear from a simple examination of the Dalitz plot shown in Figure
\ref{fig:dalitz} that there are at least three resonant 
contributions to the $D^0 \rightarrow \pi^-\pi^+\pi^0$ decay: 
$\rho^0\pi^0$, $\rho^+\pi^-$, and $\rho^-\pi^+$. Not obvious is the extent to
which other resonances may also be contributing, and the level at which a non-resonant 
component is present.  As described below, our procedure was to start with the 
three $\rho\pi$ resonances and add other components one at a time to see which 
improved the fit and yielded fit fractions which were not consistent with zero.

The long lifetime of the \ks means
that the two-body decay $D^0 \rightarrow K^0_S \pi^0$  will not
interfere with any other resonances, thus we treat the \ks
as part of the background.

\section{Analysis}
\subsection{General Overview}

We are searching for the following decay chain:
\footnote{Charge conjugation implied throughout.}
\\

{$D^{*+} \, \rightarrow \, \begin{array}[t]{l} D^0
\, \red {\bf \pi^+_{slow}} \\ \, \rightLarrow ~ \red {\bf \pi^- \,
\pi^+} \begin{array}[t]{l} \pi^0 \\ \rightLarrow ~ \red {\bf \gamma
\,\gamma } \end{array} \end{array}$}\\

We require that
the $D^0$ mesons under study come from decaying $D^{*+}$'s.  
This additional constraint greatly reduces the background without
a large signal loss, and since the charge of $\pi^+_{slow}$ is correlated to the 
charge of the charm quark in the $D$ meson, it also provide a 
method for differentiating between \dppp~ and \dbppp~ which are otherwise 
indistinguishable.

We first construct $\pi^0$ candidates from photon candidates in the
electromagnetic calorimeter.  Next, we construct $D^{0}$ candidates
by combining $\pi^0$ with pairs of oppositely charged tracks.
$D^{*+}$'s are built by adding $\pi_{slow}^+$ candidates to the $D^{0}$'s. 
Combinatoric background is significantly reduced by demanding that the mass 
of the $D$ and the $D^*-D$ mass difference both be within one 
standard deviation of the nominal value, and that the scaled momentum of the 
$D^*$, $x_{D^*} = {p_{D^*}}/{p_{D^*,max}}$, is greater than 0.7.

\subsection{Dalitz Fitter}
To fit the data to the matrix element model shown in Equation~\ref{eqn:model}, 
we used a MINUIT-based unbinned maximum likelihood fitter, minimizing

\be
\label{eqn:loglike}
{\mathcal F} = {\sum_{events}(-2 \ln {\mathcal L})},
\ee
where
\be
\label{eqn:likely}
{\mathcal L} = \left( F~ {{{\mathcal E}(m^2_{\pi^-\pi^+}, m^2_{\pi^+\pi^0})
\left| {\mathcal M}_{D^0 \rightarrow \pi^- \pi+ \pi^0} \right|^2 }
\over{ {\mathcal N}_{signal}}}
+ (1 - F) {{{\mathcal B}(m^2_{\pi^-\pi^+}, m^2_{\pi^+\pi^0})}\over{{\mathcal N}_{background}}} \right)
\ee
\vspace*{0.25in}

\noindent
represents the likelihood that a given candidate is either signal or background as a function of the
fit parameters which determine ${\mathcal M}_{D^0 \rightarrow \pi^- \pi+ \pi^0}$, and
\bi
\item $F$ is the fraction of signal events in the sample, about 80\% in this analysis,
obtained by fitting the \dppp~ mass distribution.

\item ${\mathcal E}(m^2_{\pi^-\pi^+}, m^2_{\pi^+\pi^0})$ is the
efficiency for an event falling at point $(m^2_{\pi^-\pi^+},
m^2_{\pi^+\pi^0})$ in the Dalitz plot to be detected by CLEO and to
pass all of our analysis cuts.  This shape was determined by fitting a
$2D$ cubic polynomial to reconstructed signal Monte Carlo generated uniformly
in phase-space.

\item ${\mathcal B}(m^2_{\pi^-\pi^+}, m^2_{\pi^+\pi^0})$ is the
background level at point $(m^2_{\pi^-\pi^+}, m^2_{\pi^+\pi^0})$.
The background shape was studied using data from a
$D^{*+} - D^0$ mass difference sideband, and was
parameterized by a cubic polynomial plus additional terms
representing real \ks and $\rho$ meson decays.

\item ${\mathcal N}_{signal} = {\Huge \int} {{\mathcal E}(m^2_{\pi^-\pi^+},
m^2_{\pi^+\pi^0}) \left| {\mathcal M}_{D^0 \rightarrow \pi^- \pi+
\pi^0}\right|^2} d{\mathcal DP}$ is the signal normalization.

\item ${\mathcal N}_{background} = {\Huge \int} {\mathcal B}(m^2_{\pi^-\pi^+},
m^2_{\pi^+\pi^0}) d{\mathcal DP}$ is the background normalization.

\ei

\section{Preliminary Results}

\begin{figure}[htbp]
\centerline{
\epsfxsize=2.1in\epsfbox{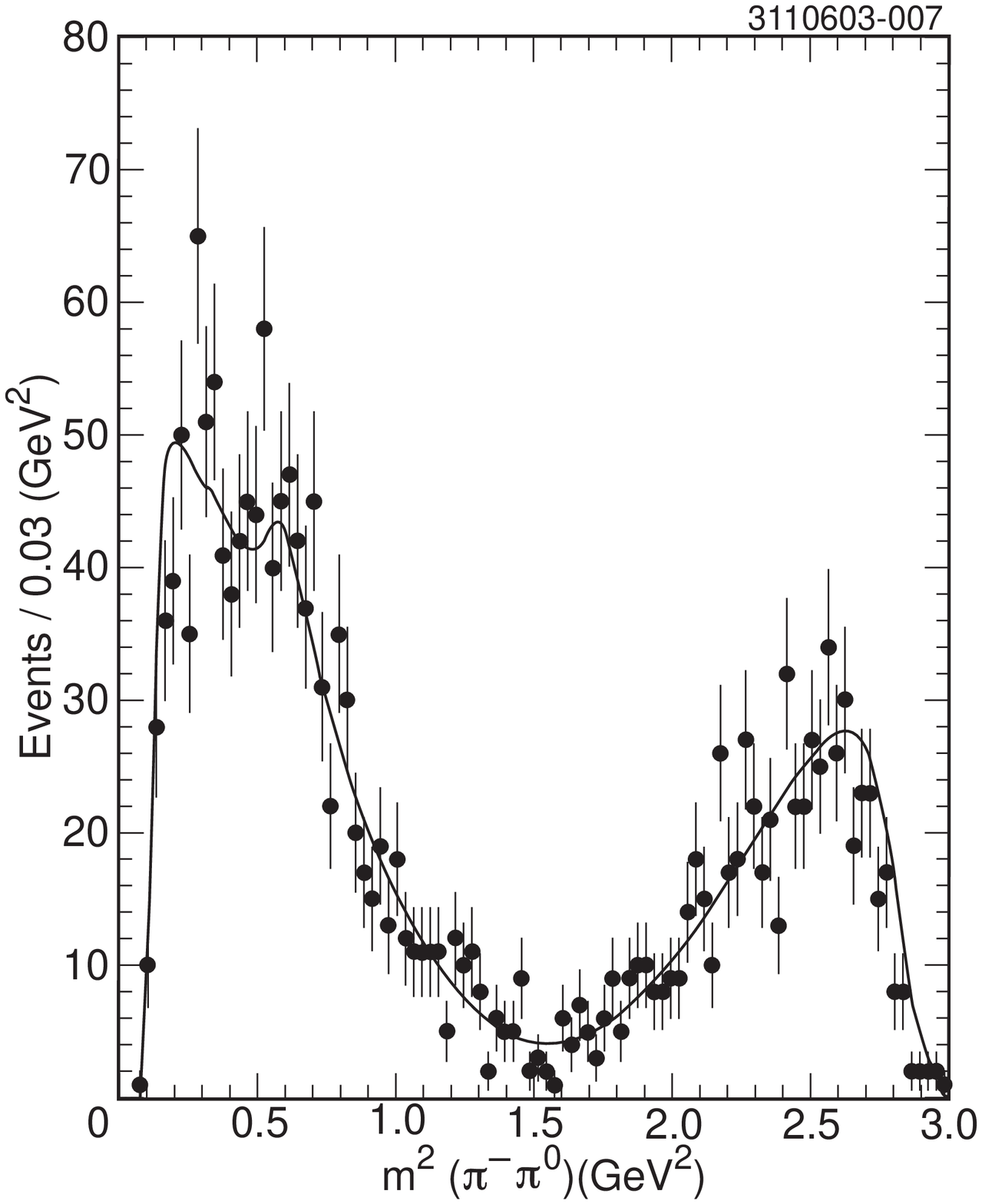}
\epsfxsize=2.1in\epsfbox{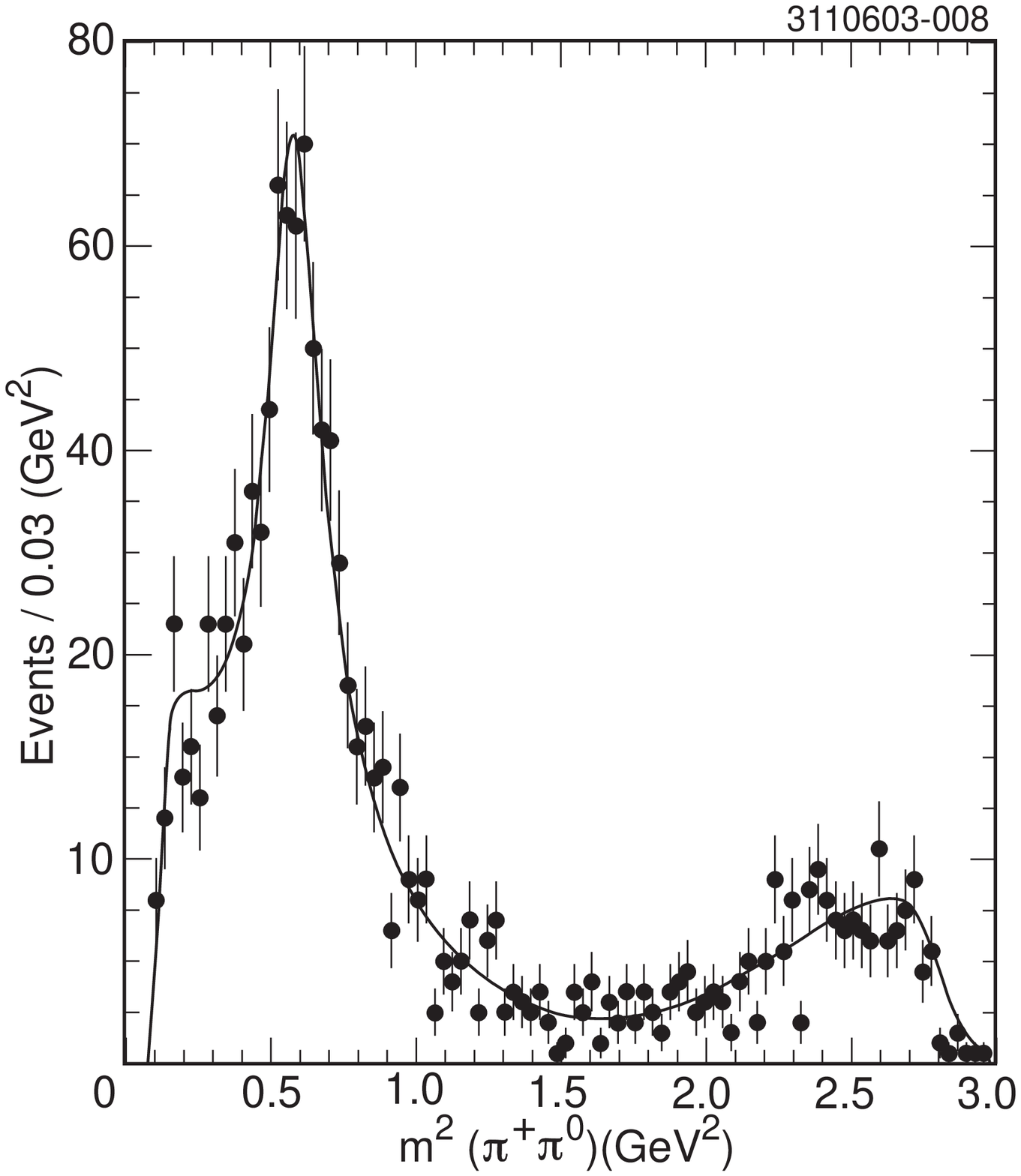}
\epsfxsize=2.1in\epsfbox{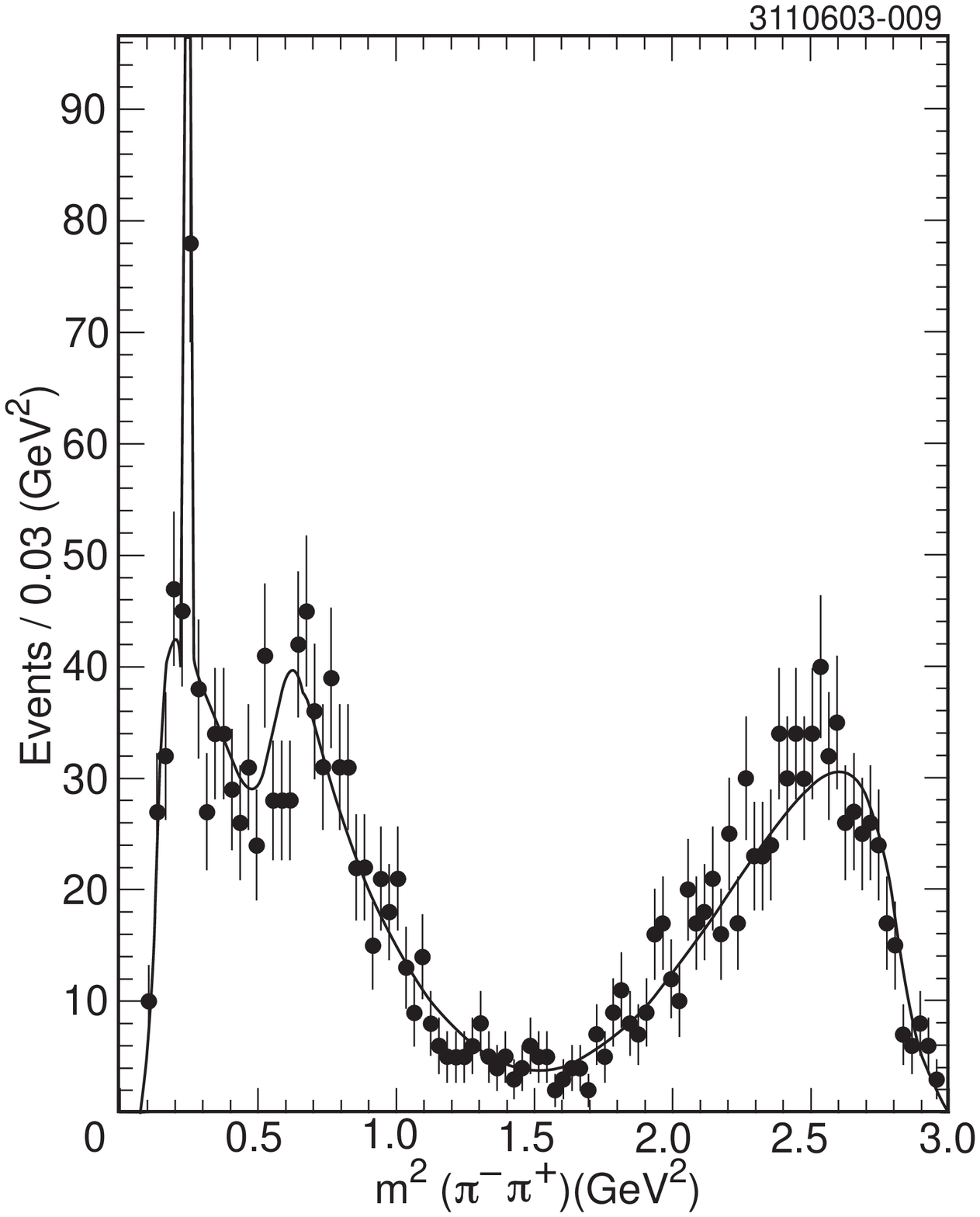}
}
\caption{ \label{fig:projections}
Projections of the $D^0 \rightarrow \pi^- \pi^+ \pi^0$ Dalitz plot on the three
mass-squared axes. The solid lines are binned projections of the best fit.}
\end{figure}

The preliminary results shown below represent the full CLEO II.V
\cite{cleoii,cleoiiv} dataset of 9 $fb^{-1}$.  
Figure~\ref{fig:projections} shows the three binned Dalitz plot projections 
for both the data and the best fit.

The only significant contribution to the resonant substructure of this decay was
seen from the  $D^0 \rightarrow \rho^+ \pi^-$, $D^0 \rightarrow
\rho^- \pi^+$, and $D^0 \rightarrow \rho^0 \pi^0$ channels. The preliminary fit
results are presented in Table~\ref{table:results}.

\btabenvh
\bc
\renewcommand{\baselinestretch}{0.9}\normalsize\small
\begin{tabular}{|c|r|r|r|}
\hline
Resonance & Amplitude & Phase($^o$) & Fit Fraction(\%)  \\
\hline
\hline

$\rho^+$    &    $1.$ (fixed)              &     $0.$ (fixed)        &   $76.5 \pm 1.8 \pm 4.8$ \\
\hline
$\rho^0$    &   $ 0.56 \pm 0.02 \pm 0.07$  &   $  10 \pm   3 \pm 3$  &   $23.9 \pm 1.8 \pm 4.6$ \\
\hline
$\rho^-$    &   $ 0.65 \pm 0.03 \pm 0.04$  &   $  -4 \pm   3 \pm 4$  &   $32.3 \pm 2.1 \pm 2.2$ \\
\hline
$non~res.$  &   $ 1.03 \pm 0.17 \pm 0.31$  &   $  77 \pm   8 \pm 11$ &   $ 2.7 \pm 0.9 \pm 1.7$ \\
\hline
\end{tabular}\\
\ec
\caption{ \label{table:results} Preliminary Fit Results}
\etabenv

Adding other resonances to the fit, including a scalar $\sigma(500)$, did not 
result in a significantly improved likelihood and yielded fit fractions 
consistent with zero.\footnote{Work to produce upper limits is under way.}

\smallskip
In addition, there was no evidence of CP asymmetry, either by comparison of the 
phases and amplitudes from separate fits to our \dppp\ and \dbppp\ data samples,  
or in the single calculated figure of merit 
${\mathcal A}_{CP} = 0.01^{+0.09}_{-0.07} \pm 0.09 $, where

\be {\mathcal A}_{CP} = {{\int {
    {\left| {\mathcal M}_{D^0} \right|^2 -
    \left| {\mathcal M}_{\overline{D^0}} \right|^2}
\over
    {\left| {\mathcal M}_{D^0} \right|^2 +
    \left| {\mathcal M}_{\overline{D^0}} \right|^2}
}
d {\mathcal DP}} \Bigg/ {\int{}  d {\mathcal DP}}}.
\ee

We estimate systematic errors by varying parameters of concern and
re-running the Dalitz fitter to assess the sensitivity of the results to
these variations.  This was done for our parameterizations of
efficiency and background, our determination of signal fraction, and
our event selection criteria.  Further systematic studies are ongoing.

\section{Conclusions}
We report preliminary results from our Dalitz analysis of \dppp.  
While we find large fit fractions for the three $\rho(770)$
channels $D^0 \rightarrow \rho^+ \pi^-$, $D^0 \rightarrow \rho^-
\pi^+$, and $D^0 \rightarrow \rho^0 \pi^0$, as well as a small but
significant non-resonant contribution, we do not find significant
contributions from any other resonances including the $\sigma(500)$.

The $\sigma(500)$ remains a controversial particle.  
E791 found strong evidence for
it in their $D^+ \rightarrow \pi^- \pi^+ \pi^+$ analysis\cite{e791}, while we 
find no need for it in $D^0 \rightarrow \pi^- \pi^+ \pi^0$. This clearly demands further study.

The lack of a contribution from more massive $\rho$ mesons is also worthy of comment.  
In CLEO's analysis
\dkpp,\cite{tjb} the $\rho^+(1700)$ had a small yet statistically 
significant fit fraction. 
Having observed $D^0 \rightarrow K^- \rho^+(1700)$, one might expect to
observe $D^0
\rightarrow \pi^- \rho^+(1700)$ as well, but we do not.  
It is interesting to note
that in the \dkpp\, analysis, the nominal peak location of the
$\rho^+(1700)$ was not on the Dalitz plot so this analysis was only
sensitive to the tail of this resonance.  In \dppp, the peak of the
$\rho(1700)$ resonances are located within the Dalitz plot,
making \dppp~ a potentially more suitable laboratory for studying these heavy
resonances.  Again, this begs for further investigation.

Finally, we saw no evidence for CP violation, either from comparing
the amplitudes, phases, and fit fractions from the separate \dppp ~and
\dbppp ~fits, or in the single \acp number.  Recent theoretical
work permits CP violation on the order of $0.1\%$
\cite{Buccella,Santorelli} in this mode, but we do not have sufficient statistics
to confront this prediction.

% CURRENT acknowledgements go here...
% download from the CLEO website 
% Hard-wired - RSG - 03Jun03
%\input{ack_prl_nov02.tex}
We gratefully acknowledge the effort of the CESR staff 
in providing us with
excellent luminosity and running conditions.
This work was supported by 
the National Science Foundation,
the U.S. Department of Energy,
the Research Corporation,
and the 
Texas Advanced Research Program.

%Example of how to insert eps files as figures
%\begin{figure}
% \includegraphics*[width=3.75in]{dspppz_dmass.eps}
% \caption{}
% \label{fig:massplots}
%\end{figure}

\end{document}